\documentclass[10pt,twocolumn,letterpaper,groupedaddress]{revtex4}

\usepackage{multibbl}
\usepackage{graphicx,amsmath,SIunits}
\usepackage{amsmath,SIunits}
\usepackage[latin1]{inputenc}
\usepackage{latexsym,stmaryrd}
\usepackage[sort&compress]{natbib}
\bibpunct{}{}{,}{s}{}{,}

\makeatletter
\newcommand*{\citenst}[2][]{%
  \begingroup
  \let\NAT@mbox=\mbox
  \let\@cite\NAT@citenum
  \let\NAT@space\NAT@spacechar
  \let\NAT@super@kern\relax
  \renewcommand\NAT@open{[}%
  \renewcommand\NAT@close{]}%
  \citep{#2}%
  \endgroup
}
\makeatother

\begin{document}

\title{Quantifying the intrinsic amount of fabrication disorder in photonic-crystal waveguides from optical far-field intensity measurements}

\author{P. D. Garc\'{i}a}
\email{garcia@nbi.ku.dk}
\author{A. Javadi}
\author{H. Thyrrestrup}
\author{P. Lodahl}
\email{lodahl@nbi.ku.dk}
\homepage{http://www.quantum-photonics.dk/}

\affiliation{Niels Bohr Institute,\ University of Copenhagen,\ Blegdamsvej 17,\ DK-2100 Copenhagen,\ Denmark}

\date{\today}

\small

\begin{abstract}
Residual disorder due to fabrication imperfections has important impact in nanophotonics where it may degrade device performance by increasing radiation loss or spontaneously trap light by Anderson localization.\ We propose and demonstrate experimentally a method of quantifying the intrinsic amount of disorder in state-of-the-art photonic-crystal waveguides from far-field measurements of the Anderson-localized modes.\ This is achieved by comparing the spectral range that Anderson localization is observed to numerical simulations and the method offers sensitivity down to $\simeq 1\,\text{nm}$.
\end{abstract}

 \pacs{(42.25.Dd, 42.25.Fx, 46.65.+g, 42.70.Qs)}

\maketitle

Optical nanostructures are able to control the emission and propagation of light at the nanoscale.\ \cite{Joannopoulos} Such structures are generally made by careful engineering where fabrication imperfections give rise to ubiquitous disorder, which may degrade the performance of nanophotonics devices inducing losses \cite{Gerace,Hughes} or Anderson localization of light.\ \cite{Anderson,Lalanne,Luca,C0} Given the maturity of current fabrication processes, quantifying the intrinsic amount of disorder in photonic nanostructures is challenging since any imaging technique would require subnanometer spatial resolution to accurately resolve the structural features.\ Important progress has been made in order to measure such unavoidable disorder by, e.g, statistical analysis of scanning electron micrographs.\ \cite{Thomas} However, reconstruction of actual dielectric profiles from scanning electron microscopy is not straightforward.\ \cite{Begin} Here we demonstrate a way of quantifying the intrinsic disorder of photonic-crystal waveguides from optical far-field intensity measurements.\ Our approach offers an accurate statistical  measurement of the imperfections with nanometer resolution.

\begin{figure}[t!]
  \includegraphics[width=8cm]{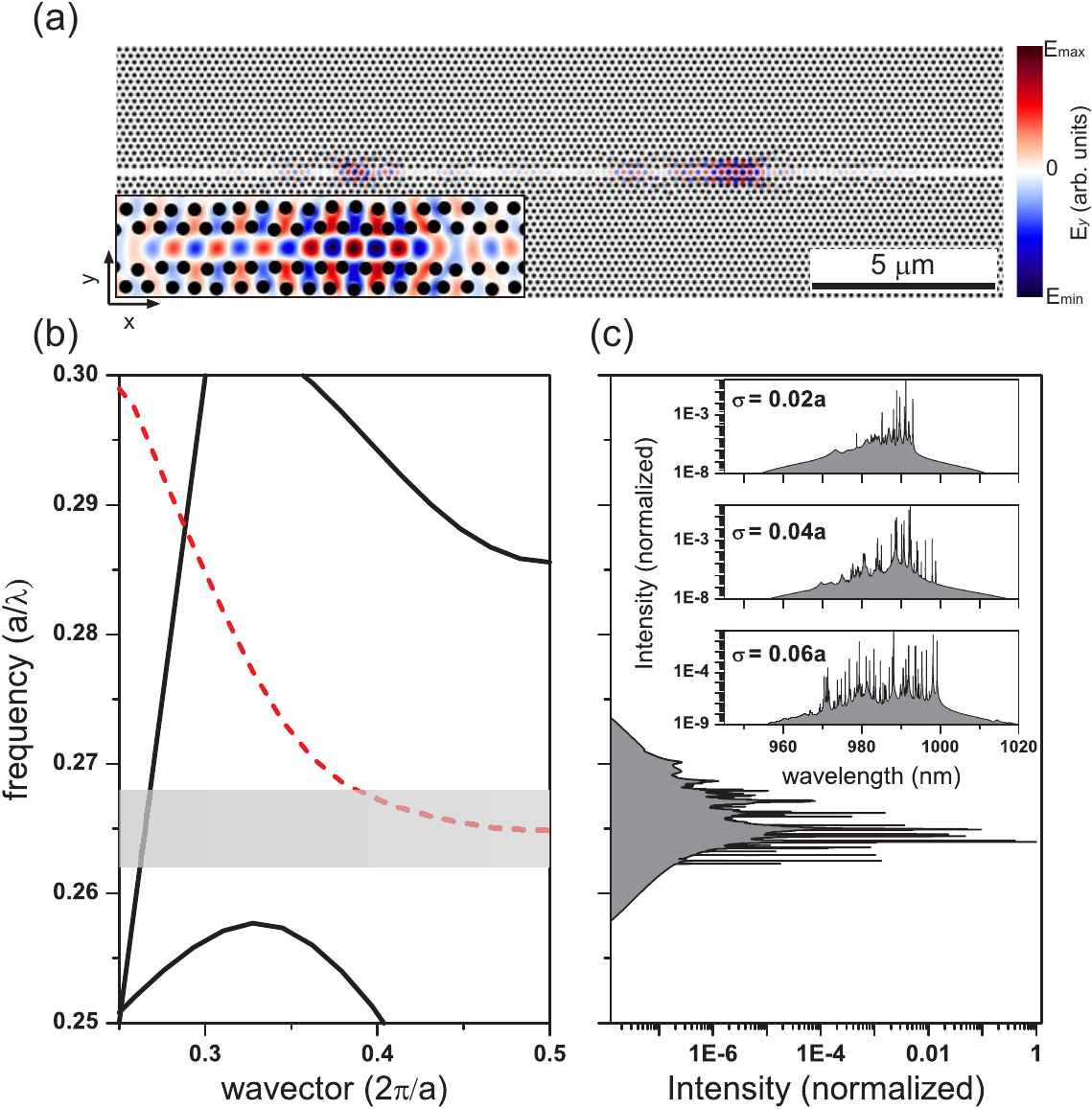}
      \caption{ \label{1} (Color online) (a) Two-dimensional finite-difference time-domain calculation of the electromagnetic field component perpendicular to a photonic-crystal waveguide (color scale).\ The parameters used for the calculation are given in the text.\ Disorder is introduced in the structure by randomly displacing the holes position in the three rows on each side of the waveguide following a normal distribution with standard deviation of $\sigma=0.04a$ (see inset).\ (b) Dispersion relation of an even-parity photonic crystal guided mode (dashed-red curve).\ The shaded region around the mode cutoff outlines the spectral region where Anderson-localized modes appear determining the Lifshitz tail.\ (c) Calculated normalized electromagnetic field intensity ensemble-averaged over 20 different positions along a single waveguide and a total of 10 disordered photonic-crystal waveguides with $\sigma=0.04a$.\ The residual speckles in the spectrum are due to the finite statistics.\ The spectral width of the region where localized modes appear can be determined precisely and used to extract the degree of disorder in experimental samples.\ The inset shows the Lifshitz tails for different amounts of disorder.}
    \end{figure}

As initially explored in the context of electronic transport, the general eigenstates of a perturbed Hamiltonian describing a crystal affected by structural disorder was obtained by I. M. Lifshitz.\ \cite{Lifshitz} The energy spectrum contains states that are different from propagating Bloch solutions and, on average, decay exponentially with a characteristic length scale called the localization length.\ These impurity levels populate the spectrum near the band edge of the unperturbed crystal with narrow resonances and smear out the density of states (DOS) by forming the so-called Lifshitz tail.\ This situation applies not only to the electronic wave case but to any kind of wave propagating through a perturbed periodic lattice.\ The optical Lifshitz tail has been probed by scanning near-field optical microscopy \cite{Huisman} and by single quantum-emitter spectroscopy.\ \cite{TyrrestrupAPL} The width of the Lifshitz tail is determined by the impurity density \cite{Lifshitz} or, in nanophotonic structures, by the amount of structural disorder.\ This dependence can be used to quantify the residual disorder of photonic nanostructures.

\begin{figure}[t!]
  \includegraphics[width=\columnwidth]{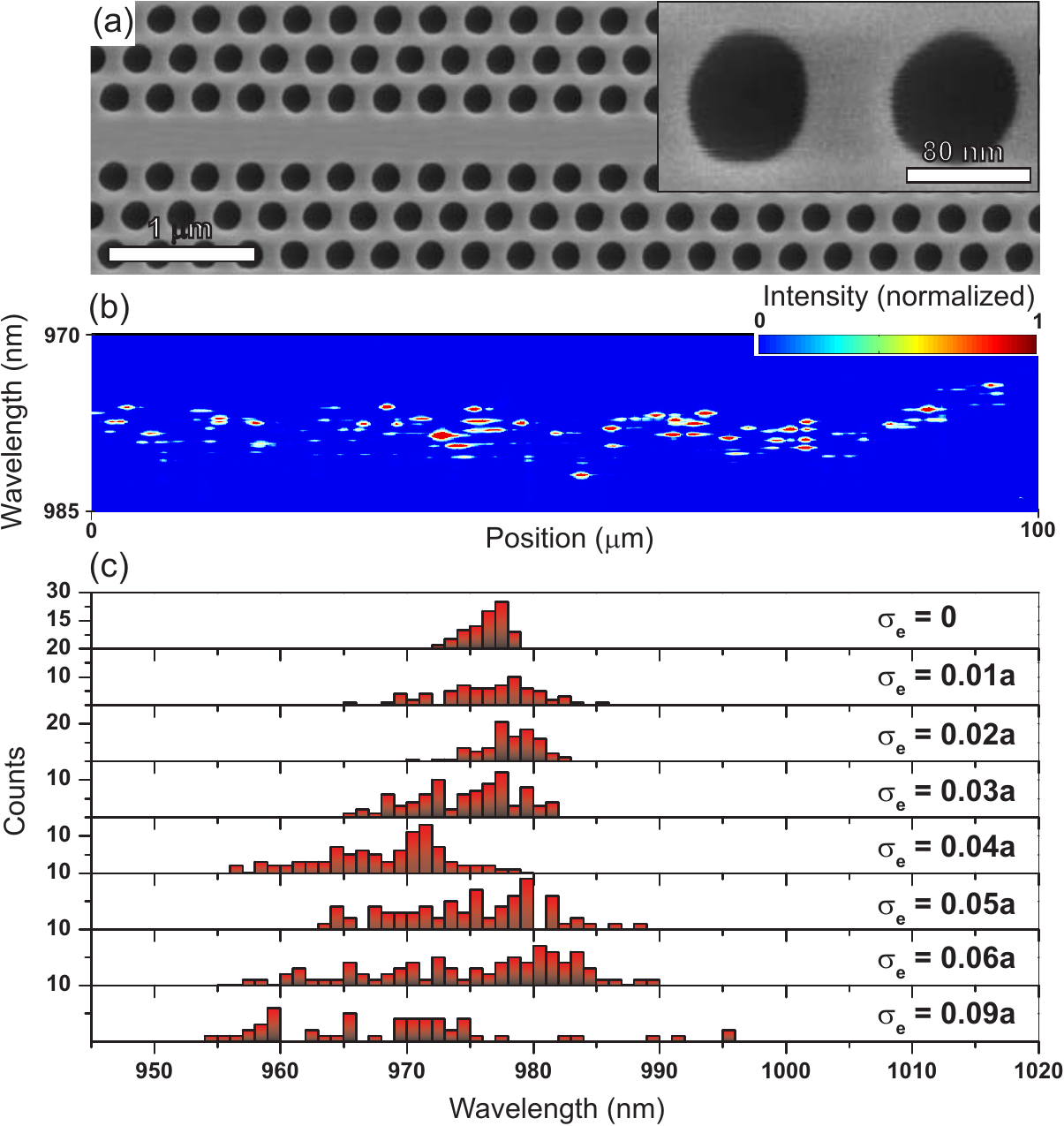}
    \caption{ \label{2} (Color online) (a) Scanning-electron micrograph of a photonic-crystal waveguide (top view) without engineered disorder.\ The inset illustrates that unavoidable irregular shapes and sizes of the holes are present in the samples.\ (b) Normalized  high-power photoluminescence spectra collected while scanning a microscope objective along a photonic-crystal waveguide with only intrinsic disorder ($\sigma_\text{e}=0$).\ The bright peaks that appear randomly along the waveguide and close to the waveguide mode cutoff correspond to Anderson-localized modes.\ In order to display the peaks of varying heights, the plot is saturated at $10\,\%$ of the maximum peak intensity.\ The spectral region where the modes appear defines the experimental Lifshitz tail.\ (c) Histogram of the spectral position of the Anderson-localized modes for different amounts of engineered disorder, $\sigma_\text{e}$.\ The width of the Lifshitz tail increases when increasing the amount of disorder.}
    \end{figure}

We use two-dimensional finite-difference time-domain (FDTD) simulations \cite{fdtd} to calculate the Lifshitz tail of photonic-crystal waveguides using a freely available software package.\ \cite{meep} The simulated structures replicate the fabricated ones and consist of a hexagonal lattice of air holes forming a two-dimensional photonic crystal with a lattice constant $a=260\,\text{nm}$ and a hole radius $0.29a$.\ A waveguide is introduced in this structure by leaving out a row of holes.\ The length of the simulation domain is $120a$, with eight rows of holes on each side of the waveguide.\ The position of the holes in the three rows on both sides of the waveguide are randomly displaced a certain amount, $\Delta \text{\textbf{r}}$, normally distributed with a standard deviation $\sigma = \sqrt{<\Delta \text{\textbf{r}}^2> - <\Delta \text{\textbf{r}}>^2}$, where the brackets indicate the expectation value.\  By using an effective refractive index  of $n=2.76$, the $150\,\micro\text{m}$ thick photonic-crystal membrane can effectively be simulated in 2D, \cite{effective_refractive_index} which significantly improves computation times.\ Perfectly matching layers (PML) cover 10 lattice constants at both waveguide terminations in order to model an open system.\ \cite{meep} A dramatic effect of disorder in a photonic-crystal waveguide is one-dimensional Anderson localization, which confines light randomly along the waveguide forming a narrow spectral band of strongly localized resonances around the waveguide cutoff frequency.\ \cite{Vollmer,Luca} We use a total of nine electric dipole sources randomly positioned along the waveguide to excite the localized modes in a broad bandwidth of $\Delta \omega = 0.02 (a/\lambda)$, centered at the waveguide mode cutoff frequency.\ From the resulting electric field, $\textbf{E}(\textbf{r},t)$, we can calculate the intensity spectrum, $I(\omega) = |\mathcal{FFT}[\textbf{E}(\textbf{r},t)] |^2$ that consists of a set of Lorentzian peaks due to the Anderson-localized quasimodes.\ To circumvent the limited spectral resolution of the Fourier transform, we use a method denoted harmonic inversion, which allows resolving spectrally-close resonances and to extract their Q-factors with high accuracy.\ \cite{Harminv} Further details of the simulations can be found in Ref.~\citenst{TyrrestrupPHD}.\ Figure~\ref{1}(a) plots the in-plane component of the electric field perpendicular to the photonic-crystal waveguide, $\textbf{E}_{y}(\textbf{r})$, revealing strongly-confined modes randomly appearing along the waveguide.

Figure~\ref{1}(b) shows the dispersion relation, $\omega(k)$, of the unperturbed structure, \cite{mpb} where $k$ is the length of the wave vector along the waveguide.\ The fundamental TE-guided mode (dashed-red line) gives rise to a strongly increasing DOS $(\text{DOS}=(1/\pi) \partial k/\partial \omega)$, enabling slow light with very low group velocity $(v_\text{g} \propto 1/\text{DOS})$ \cite{Baba} and efficient broadband single-photon sources.\ \cite{Toke,TyrrestrupAPL} Residual disorder smears the DOS \cite{Savona_PRB}, sets a limitation to slow light, and may induce localized modes in the slow-light regime if the sample length and the characteristic length related to out-of-plane losses exceed the localization length.\ \cite{Garcia2010,Smolka2011} Figure~\ref{1}(c) shows the calculated ensemble-averaged intensity, $I(\omega)$, displaying the spectral range of the Lifshitz tail.\ The observed sharp resonances illustrate that the finite statistics implies that the ensemble average is not fully converged, but nonetheless the width of the Lifshitz tail can be precisely extracted.\ We define the width of the Lifshitz tail, $\Delta \omega$, as the spectral range where sharp resonances appear, i.e., peaks that are, at least, 10 times brighter than at neighboring frequencies.\  As shown in the inset of Fig.~\ref{1}(c), the width of the Lifshitz tail broadens when increasing $\sigma$ and we can use this dependence to extract the intrinsic amount of disorder in the photonic-crystal waveguides.

Lifshitz tails are measured in $100\,\micro\text{m}$ long GaAs photonic-crystal waveguide membranes containing a single layer of self-assembled InAs quantum dots (QDs) in the center and otherwise identical parameters as the simulated structures (see Fig.~\ref{2}(a), details of the samples can be found in Ref.~\citenst{Luca}).\ Due to the tolerance of the fabrication process, imperfections in the shape, size, and position of the holes are unavoidable, see inset of Fig.~\ref{2}(a).\ A controlled degree of disorder in the hole position is introduced in the three rows of holes on both sides of the waveguide in the same way as discussed for the simulations.\ The embedded QDs are used to excite the Anderson-localized modes by using a confocal micro-photoluminescence setup for excitation and collection~\citenst{Toke,Smolka2011}.\ Photoluminescence emission spectra are collected in a wide wavelength range of $900\,\text{nm} < \lambda < 1000\,\text{nm}$ while scanning the microscope objective along the photonic-crystal waveguides, as shown in Fig.~\ref{2}(b).\ Clear Anderson-localized modes are observed within a narrow spectral range around the waveguide cutoff ($\lambda = 978\,\text{nm}$), which provides the width of the Lifshitz tail.\ Figure~\ref{2}(c) displays the histogram of the Anderson-localized modes spectral positions for different amounts of engineered disorder, showing how the Lifshitz tail broadens spectrally when increasing the amount of disorder as expected from the simulations.

\begin{figure}[t!]
  \includegraphics[width=6cm]{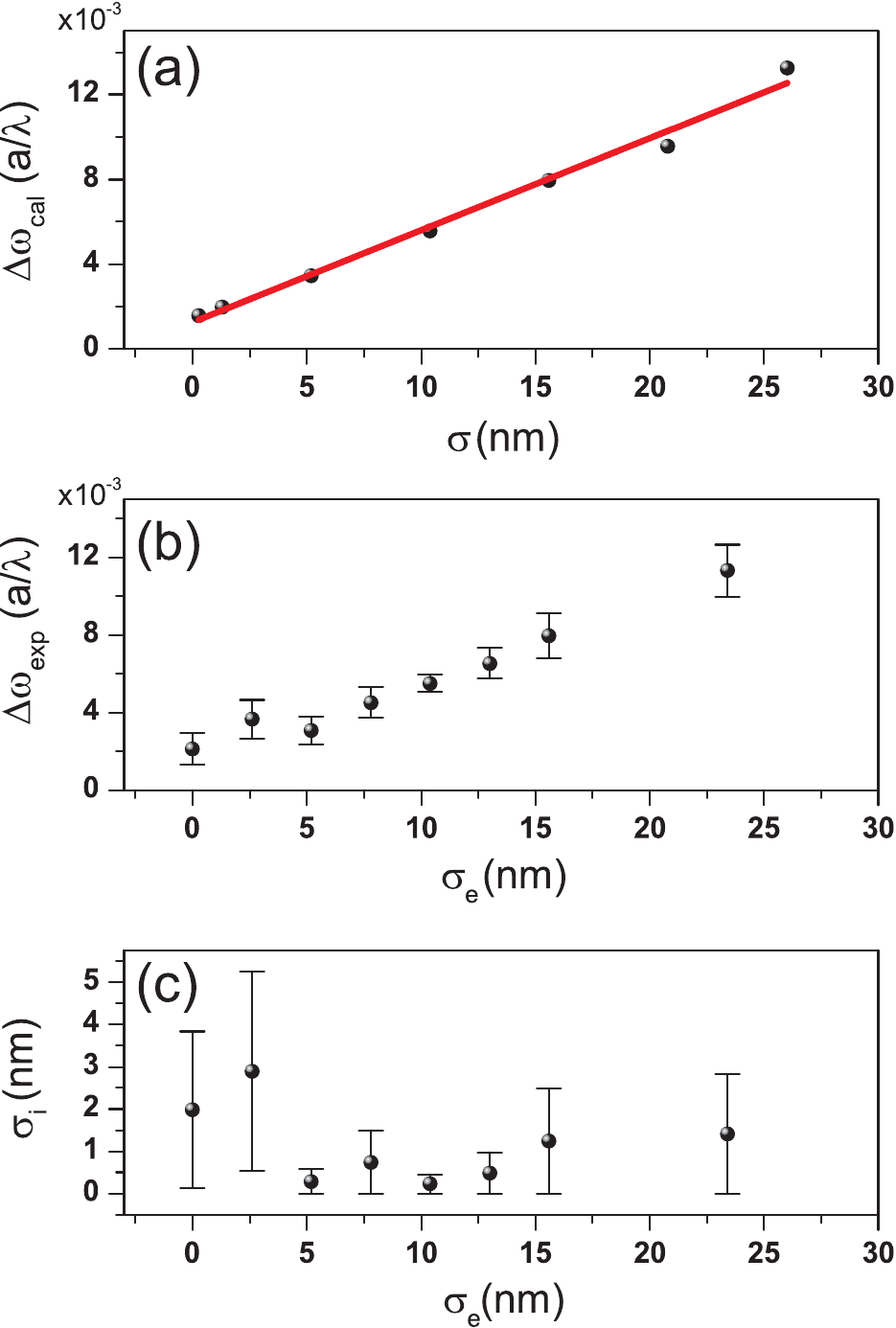}
    \caption{ \label{3} (Color online) (a) Calculated width of the Lifshitz tail for photonic-crystal waveguides with a different standard deviation in the hole positions and $a=260\,\text{nm}$.\ The fit to a linear dependence on disorder (red line) is very satisfactory for small amounts of disorder ($ \leq \sigma = 0.1a$).\ (b) Experimental width of the Lifshitz tail measured for different amounts of engineered disorder.\ (c) Recorded intrinsic degree of disorder for samples with different engineered disorder.}
\end{figure}

For small amounts of disorder, the calculated width of the Lifshitz tail depends linearly on the amount of disorder, i.e., $\Delta \omega_\text{cal} \propto \sigma$, as displayed in Fig.~\ref{3}a.\ Note that in the limit of very small amount of disorder the applied boundary conditions do not suppress stray reflections at the end of the photonic-crystal waveguide sufficiently well, which appears as a finite width of the Lifshitz tail.\ As predicted by the simulations, the width of the Lifshitz tail recorded experimentally, $\Delta \omega_\text{exp}$, is found to vary linearly with the engineered disorder, see Fig.~\ref{3}b.\ It turns out that the Anderson-localized modes appearing close to the waveguide terminations are spectrally shifted.\ In order to avoid such finite-size effects, the width of the Lifshitz tail is calculated by dividing the waveguide into sections of $20\,\micro\text{m}$ length and integrate the spectrum over each of these sections.\ The mean value and standard deviation of the Lifshitz tail width  are shown in Fig.~\ref{3}b for different amounts of disorder.\ These data can be used to extract the intrinsic disorder associated with the fabrication process. Thus, the total degree of disorder in the samples $\sigma$ consists of both intrinsic ($\sigma_i$) and engineered disorder ($\sigma_e$) that are assumed to be independent and therefore related through

\begin{equation}\label{variance}
   \sigma^2 = \sigma_\text{i}^2  + \sigma_\text{e}^2.
\end{equation}
Making use of the predicted linear relation between $\Delta \omega$ and $\sigma$ enables us therefore to extract $\sigma_i$ from the data in Fig.~\ref{3}b.\ It is displayed in Fig.~\ref{3}c for the various samples with different amounts of engineered disorder, and as anticipated no correlation between the two is observed.\ While the method does not identify the nature of the intrinsic  disorder, we are able to extract quantitative numbers for the statistical properties of the disorder. Thus the average degree of intrinsic disorder found in the samples is equivalent to hole displacements of $<\sigma_\text{i}>=(1.2 \pm 1.0)\text{nm}$.\ These results appear to be comparable to the numbers  inferred from optical near-field measurements of the Lifshitz tail \cite{Huisman} and from statistical analysis of scanning electron micrographs \cite{Thomas}.\ We emphasize the remarkable sensitivity of the applied method since it allows to probe disorder at the sub-nanometer length scale. Furthermore, using the QDs as internal emitters for exciting the Anderson-localized modes appears a very attractive way of quantifying disorder  compared to tedious statistical analysis or complicated phase-sensitive optical near-field measurements.

In conclusion, we have measured the spectral width of the Lifshitz tail of photonic-crystal waveguides with different amounts of engineered disorder in the regime of 1D Anderson localization.\ By comparing the experimental data to numerical simulations, we are able to quantify the intrinsic disorder of the samples with nanometer resolution.\ The method proposed here could potentially be implemented in other disordered nanophotonic structures as well in order to quantify and potentially optimize the degree of disorder within the setting of cavity quantum-electrodynamics \cite{Tyrrestrup} or random lasing \cite{Cao,Jin}.

\textbf{Acknowledgements}

We gratefully acknowledge financial support from the Danish council for independent
research (natural sciences and technology and production sciences), and the European research council (ERC consolidator grant "ALLQUANTUM").


\end{document}